\documentclass[%
nofootinbib,
 amsmath,amssymb,
 aps,longbibliography,twocolumn
]{revtex4-1}

\usepackage{graphicx}
\usepackage{bm}
\usepackage{gensymb}
\usepackage{mathtools}
\usepackage{lineno}
\usepackage[ colorlinks = true,              linkcolor = blue, urlcolor  =
blue,              citecolor = red,              anchorcolor = green,
]{hyperref}

\usepackage{newtxtext,newtxmath}

\providecommand{\U}[1]{\protect\rule{.1in}{.1in}}

\numberwithin{theorem}{section}
\numberwithin{proposition}{section}
\numberwithin{definition}{section}
\numberwithin{example}{section}

\newcommand{\bra}[1]{\langle#1|}
\newcommand{\ket}[1]{|#1\rangle}

\allowdisplaybreaks

\begin{document}

\preprint{APS/123-QED}

\title{Quantum Algorithms for Testing Hamiltonian Symmetry}

\author{Margarite L. LaBorde}
\author{Mark M. Wilde}
\affiliation{
 Hearne Institute for Theoretical Physics, Department of Physics and Astronomy,
and Center for Computation and Technology, Louisiana State University, Baton
Rouge, Louisiana 70803, USA
}

\date{\today}

\begin{abstract}
Symmetries in a Hamiltonian play an important role in quantum physics because they correspond directly with conserved quantities of the related system. In this paper, we propose quantum algorithms capable of testing whether a Hamiltonian exhibits symmetry with respect to a group. We demonstrate that familiar expressions of Hamiltonian symmetry in quantum mechanics correspond directly with the acceptance probabilities of our algorithms. We execute one of our symmetry-testing algorithms on
existing quantum computers for simple examples of both symmetric and asymmetric cases.
\end{abstract}

\maketitle

\textit{\label{sec:level1}Introduction}---Symmetry is a key facet of nature that plays a fundamental role in physics \cite{Gross96,FR96}. Noether's theorem states that symmetries in Hamiltonians correspond with conserved quantities in the related physical systems \cite{Noether1918}. The symmetries of a Hamiltonian indicate the presence of superselection rules \cite{PhysRev.155.1428, Wick1952}. In quantum computing and information, symmetry can indicate the presence of resources or lack thereof \cite{marvian2012symmetry}, and it can be useful for improving the performance of variational quantum algorithms \cite{SSY20,Gard2020,BGAMBE21,LXYB22}. Identification of symmetries can simplify calculations by eliminating degrees of freedom associated with conserved quantities---this is at the heart of Noether's theorem. This makes symmetries extraordinarily useful in the context of physics.

Quantum computing is a significantly younger field of study. First introduced as a quantum-mechanical model of a Turing machine \cite{benioff1980computer}, the intrigue of quantum computers lies in their potential to outperform their classical counterparts. 
The most obvious asset of quantum computers is the inherent physics behind the calculation, which includes non-classical features such as superposition and entanglement. Classical simulations of quantum systems quickly become intractable as the size of the Hilbert space grows, needing exponentially many bits to explore the state space which multiple qubits naturally occupy. Intuitively, the quantum mechanical nature of these computers allows for simulations of quantum systems in a forthright way (see \cite{childs2018toward} and references therein). 

A pertinent example of this, Hamiltonian simulation \cite{lloyd1996universal}, garners high interest in the field \cite{blatt2012quantum,somma2016trotter,cubitt2018universal,clinton2021hamiltonian}. Much work has been done to understand how to simulate these dynamics on quantum hardware such that they can be efficiently realized; however, to the best of our knowledge, there is currently no algorithm that tests Hamiltonian symmetries on a quantum computer, even though simulating Hamiltonians in this manner and identifying the symmetries of said Hamiltonian are both deemed to be of utmost importance.

In this paper, we give quantum algorithms to test whether a Hamiltonian evolution is symmetric with respect to the action of a discrete, finite group. This property is often referred to as the covariance \cite{CDP09} of the evolution. If the evolution is symmetric, then the Hamiltonian itself is also symmetric, and so our algorithms thus tests for Hamiltonian symmetry. Furthermore, we show that for a Hamiltonian with an efficiently realizable unitary evolution, we can perform our first test efficiently on a quantum computer \cite{clinton2021hamiltonian}. ``Efficiently'' here means that the time necessary to complete the calculation to within a constant error bound scales at most polynomially with the number of qubits in the system. Our second quantum algorithm for testing Hamiltonian symmetry can be implemented by means of a variational approach \cite{CABBEFMMYCC20,bharti2021noisy}. The acceptance probabilities of both algorithms can be elegantly expressed in terms of familiar expressions of Hamiltonian symmetry (see Eqs.~\eqref{eq:accept-prob-comm}, \eqref{eq:accept-prob-optimized}, and \eqref{eq:lower-bnd-est-small-t}).  We note here that our algorithms can be understood as particular kinds of property tests \cite{MW16} of quantum systems. As examples, we consider the transverse-field Ising model, the Heisenberg~XY model \cite{LSM61}, and the weakly $J$-coupled NMR Hamiltonian \cite{van1996multidimensional},  whose evolution we test for various symmetry cases. 

The consequences of such results extend throughout many areas of physics. Any study of a physical Hamiltonian can benefit from finding its symmetries, and our algorithms allow for an efficient check for these symmetries. With this knowledge, dynamics can be simplified by excluding symmetry-breaking transitions, calculations can be reduced into fewer dimensions, and intuition can be gained about the system of interest. Our first algorithm also scales well, meaning that systems too large and cumbersome to be studied by hand or classical computation can instead be investigated in a practical time scale. Our quantum tests offer meaningful insight into physical dynamics.

In what follows, we begin by describing covariance symmetry of a unitary quantum channel---of which Hamiltonian dynamics are a special case.
Next, we briefly review how Hamiltonian dynamics can be simulated on a quantum computer through the Trotter--Suzuki approximation \cite{suzuki1976generalized}.  
We then present our main result---quantum algorithms to test the covariance symmetry of Hamiltonian dynamics. 
Finally, we demonstrate examples of symmetry tests on currently available quantum computers, and we discuss additional implications of our work.

\textit{\label{sec:level2} Covariance of a Quantum Channel}---Before describing the symmetries of a Hamiltonian, 
we first address the notion of covariance symmetry of a quantum channel \cite{Hol02}. Quantum channels transform one quantum state to another and are described by completely positive, trace-preserving maps. They serve as a convenient mathematical description of the dynamics induced by a Hamiltonian. The symmetries of a Hamiltonian naturally correspond to a covariance symmetry in the channel given by its evolution, and we exploit this in our algorithms.

We recall the established concept of covariance symmetry in more detail in Appendix~\ref{app:covariance}
but briefly summarize the notion here. Suppose there is a channel sending Alice's quantum system to Bob's. For simplicity, we consider their systems to have the same dimension, though this is not required in general. Further suppose that we wish to determine if this channel is symmetric with respect to some finite, discrete group $G$, which has a projective unitary representation (usually denoted $\{U(g)\}_{g\in G}$). 
Then the channel is covariant if Alice acting with her representation $U(g)$ before sending the system through the channel is completely equivalent to Bob acting on his system with his representation of $g$ after the state has been sent through the channel. In this sense, the channel commutes with the action of the group.

One method for testing this property given some channel involves using its Choi state, formally defined in Appendix~\ref{app:covariance} 
. The Choi state is generated by sending one half of a maximally-entangled state through the channel, which we now assume to be unitary. Given the same group and its unitary representation, we define a projector
\begin{equation}\label{eq:projector}
    \Pi^G \coloneqq\frac{1}{|G|} \sum_{g \in G} \overline{U}_R(g) \otimes U_B (g) ,
\end{equation}
onto the space of states of a composite system $RB$ that are symmetric with respect to the group $G$, where the overline denotes complex conjugation. (Here we use $R$ to refer to a reference system and $B$ to refer to Bob's system after the channel, a notion we use throughout.) The Choi state of the channel is equal to its projection onto the symmetric space if and only if the Choi state is symmetric with respect to $G$, given unitary representations of the system. If the Choi state of a channel exhibits this symmetry, then the channel itself is covariant \cite{CDP09}, and the converse is true as well\footnote{ This symmetry is necessarily dependent on the unitary representations used, although this is typically suppressed when referenced in the literature. We will also suppress this on the assumption that all representations are faithful.}.

This last notion of symmetry allows us to directly prescribe an algorithm to test for Hamiltonian symmetries. If we can emulate the dynamics of a Hamiltonian efficiently, we can test for the symmetry of its Choi state. The symmetry of the Choi state then directly implies symmetry of the Hamiltonian being tested.

\textit{\label{sec:level4}Quantum Simulations of Hamiltonians}---Quantum simulations provide a method for implementing Hamiltonian dynamics on quantum computers, usually by approximating them as sequences of quantum logic gates \cite{lloyd1996universal,childs2018toward}. Much work has been conducted in this field, including work on implementations on near-term hardware \cite{clinton2021hamiltonian,CCHCCS20}, simulation by qubitization \cite{low2019hamiltonian}, simulation of operator spread \cite{geller2021quantum}, and more. Here, we review an example implementation. 

One common approach \cite{lloyd1996universal} employs the Trotter--Suzuki approximation \cite{trotter1959product,suzuki1976generalized}. This method allows for decomposition into local Hamiltonian evolutions with some specified error. In this approximation, we suppose that the Hamiltonian $H$ is of the form
$ H = \sum_{i=1}^m H_i $,
where each $H_i$ is a local Hamiltonian. Then we can describe its evolution by \cite[Proposition~F.3]{childs2018toward}
\begin{equation}
\label{eq:trotter}
    e^{-i H t} = \left(\prod_{j=1}^m e^{-i H_j t/r}\right)^r + \mathcal{O}\!\left (\frac{m^2 t^2}{r}\right ) \, ,
\end{equation}
where the correction term is negligible for $m^2t^2/r \ll 1$ and vanishes when the terms in the decomposition commute. (Here and throughout, we take $\hbar=1$.) By other methods, the error can be reduced to higher orders in~$t$~\cite{BACS07}.

\textit{\label{sec:level3}An Efficient Quantum Algorithm to Test Hamiltonian Symmetries}---Given the notion of covariance recalled above and a way to simulate the applicable Hamiltonian, we now propose a quantum algorithm to test a Hamiltonian for covariance symmetry. We begin by supposing that we have a Hamiltonian composed of a finite sum of $k$-local Hamiltonians, as described previously, with dynamics realized by higher-order methods such that the simulation error is $\mathcal{O}(t^4)$. 
Then we claim a test for symmetries of this Hamiltonian with respect to a group $G$ with a projective unitary representation $\{U(g)\}_{g\in G}$ can be performed efficiently on a quantum computer.

\begin{figure}[t]
\begin{center}
\includegraphics[width=3.5in]{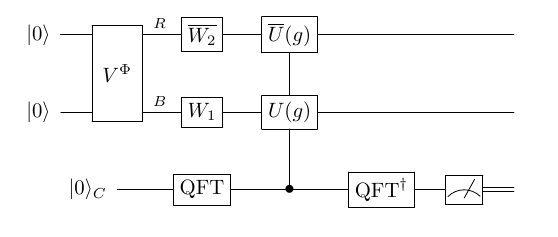}
\end{center}
\caption{Quantum circuit to test for the covariance of a unitary Hamiltonian evolution. The unitary $V^\Phi$ generates the state $\ket{\Phi}_{RA}$, the maximally-entangled state on $RA$. The evolution of the system is given by $e^{-i H t} = W_1 W_2^\dag$ and the $U(g)$ gates are controlled on a superposition over all of the elements $g \in G$, as in \eqref{eq:superpose-gr-elems}.}
\label{fig:originalcircuit}
\end{figure}

The circuit presented in Figure~\ref{fig:originalcircuit} implements such a test, and we sketch its action here. Let the input state to the circuit be the maximally-entangled state $\Phi_{RA}$. Then act on the $A$ subsystem with the unitary Hamiltonian dynamics. 
As indicated in Figure~\ref{fig:originalcircuit}, the depth of the circuit to realize this algorithm can be cut in half by taking advantage of the transpose trick $(X \otimes I) \ket{\Phi} = (I\otimes X^T)\ket{\Phi}$ and the decomposition $e^{-i H t} = W_1 W_2^\dag$, which is clearly possible for Hamiltonian simulations of the form in~\eqref{eq:trotter} or from \cite{BACS07}.  The state of the system is now given by
\begin{equation}
    \Phi_{RB}^{t} \coloneqq (\mathbb{I}_R \otimes e^{-i H t} ) \Phi_{RA} (\mathbb{I}_R \otimes e^{i H t}),
\end{equation}
which is exactly the Choi state of the channel generated by~$e^{-i H t}$. 
We then use the quantum Fourier transform (QFT) to generate a control register in the following superposed state:
\begin{equation}
    \ket{+}_C \coloneqq \frac{1}{\sqrt{|G|}}\sum_{g\in G}\ket{g}.
    \label{eq:superpose-gr-elems}
\end{equation}
Implementing the controlled  $\overline{U}(g)$ and $U(g)$ gates using the above control register yields the state
\begin{equation}
    \frac{1}{|G|}\sum_{g,g'\in G}(\overline{U}_R(g) \otimes U_B(g)) (\Phi_{RB}^{t}\otimes \ket{g} \! \bra{g'}_C) (\overline{U}^\dag_R(g') \otimes U^\dag_B(g')).
\end{equation}
Finally, we perform the measurement $\mathcal{M}=\{\ket{+}\!\bra{+}_C,\mathbb{I}-\ket{+}\!\bra{+}_C\}$ on the control register and accept if and only if the outcome $\ket{+}\!\bra{+}_C$ is observed. With this condition, the acceptance probability is given by
\begin{align}
     P_{\text{acc}} &= \operatorname{Tr}[\Pi^G \Phi_{RB}^{t}], \label{eq:originalacc}
\end{align}
where we have used the projector defined in \eqref{eq:projector} (see Appendix~\ref{app:accept} 
for a quick derivation of \eqref{eq:originalacc}). 
As a limiting case of the gentle measurement lemma \cite{Davies1969,itit1999winter,ON07}, we have that
\begin{equation}
\operatorname{Tr}[\Pi^{G}\Phi_{RB}^{t}]=1\quad\Leftrightarrow\quad \Phi_{RB}^{t}
=\Pi^{G}\Phi_{RB}^{t}\Pi^{G}, \label{eq:Bose-symmetric-equiv-cond}
\end{equation}
where the second statement is equivalent to the condition on the Choi state given in Appendix~\ref{app:covariance}. 
 Therefore, by implementing this algorithm, we can determine whether a Hamiltonian exhibits a symmetry under a group $G$ with some projective unitary representation $\{U(g)\}_{g\in G}$. See Appendix~\ref{app:approximate}  
for further details of  approximate versions of the equivalence in~\eqref{eq:Bose-symmetric-equiv-cond}, which demonstrate that the acceptance probability is near to one if and only if the Choi state is approximately Bose symmetric. 

\begin{figure}[t!]
\begin{center}
\includegraphics[
width=\linewidth
]{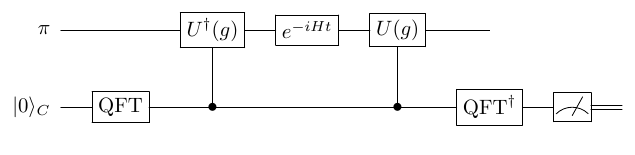}
\end{center}
\caption{Quantum circuit to test for the covariance of a unitary Hamiltonian evolution. Here, $\pi$ denotes the maximally-mixed state $\mathbb{I}/d$.
}
\label{fig:altcircuit}
\end{figure}

This algorithm can be further simplified. By invoking the transpose trick (see, e.g., \cite{biamonte2017tensor}), we can identify the unitary on the reference system, $\overline{U}_R(g)$, with an equivalent action on $A$ given by $U^\dag_A(g)$. Since the action of the circuit would then take place solely on the subsystem $A$, the reference system $R$ is traced out. This is equivalent to preparing the maximally-mixed state (denoted by $\pi$) on $A$, such that this variation of our algorithm bears some resemblance to a one-clean-qubit algorithm \cite{KL98} (also known as a DQC1 algorithm), with the exception that it requires $\log_2 |G|$ clean qubits for the control register. This simplification is shown in Figure~\ref{fig:altcircuit}. 
The acceptance probability of the simplification described above is given by
\begin{equation}
\label{eq:modifiedacc}
    P_{\text{acc}}=\frac{1}{d |G|}\sum_{g \in G}\operatorname{Tr}[U^\dag (g) e^{i H t} U(g) e^{-i H t}],
\end{equation}
where $d$ is the dimension of the system being tested. Appendix~\ref{app:accept} 
gives a proof that the expression in \eqref{eq:modifiedacc}  is equal to the acceptance probability of the circuit in Figure~\ref{fig:originalcircuit}.

The proposed circuit is limited in complexity only by the implementation of the Hamiltonian and unitary representation. Thus, our first quantum algorithm is efficiently realizable. Furthermore, we have shown that entanglement resources usually necessary for characterizing the Choi operator of a quantum channel are not necessary here. We also note that the statistics accumulated for the maximally-mixed state can be equivalently found in a sampling manner using computational basis state inputs.

We note that the acceptance probability given in \eqref{eq:modifiedacc} bears some resemblance to a group-averaged out-of-time-order correlator (OTOC) \cite{de2019spectral,swingle2016measuring,hashimoto2017out}, a measure of near-time quantum chaos. Previous work gave an efficient quantum algorithm for estimating an OTOC  \cite{pg2021exponential}; however, their work did not consider symmetry transformations of Hamiltonian evolutions nor have the group-symmetric structure considered here. Additionally, a continuous group-averaged OTOC was shown to relate to the spectral form factor \cite{de2019spectral}, a measure of late-time chaos in a system. However, it is unclear how this quantity would be interpreted for a discrete group rather than a continuous group such as previously investigated.

To provide evidence that our algorithm cannot generally be simulated efficiently by classical computers, we turn to established notions of computational complexity. In Appendix~\ref{sec:DQC1-completeness}
, we prove that estimating the acceptance probability in \eqref{eq:modifiedacc} to within additive error is a DQC1-complete problem. This means that \eqref{eq:modifiedacc} can be estimated within this restricted model of quantum computing (via our algorithm and by an observation of \cite[Section~1]{SJ08}). Furthermore, this demonstrates that estimating~\eqref{eq:modifiedacc} is just as computationally hard as any problem in this complexity class. Strong evidence exists that classical computers cannot solve DQC1-complete problems efficiently \cite{MFF14,FKMNTT18}, thus ruling out any possibility of estimating the acceptance probability in \eqref{eq:modifiedacc} by a classical sampling approach. See Appendix~\ref{sec:DQC1-completeness} 
  for further details and discussions.

\textit{A Derivation of Symmetry in the Acceptance Probability}---From the acceptance probability given in \eqref{eq:modifiedacc}, we can derive a relationship with the familiar expression of Hamiltonian symmetry in quantum mechanics, further establishing this as an authentic test of symmetry. 
Consider expanding $e^{i H t}$, under the assumption that $\tau \coloneqq \left\Vert H \right\Vert_\infty t < 1$, where $\left\Vert X\right\Vert_\infty \coloneqq \sup_{\ket{\psi} \neq 0} \frac{\left\Vert X \ket{\psi}\right\Vert_2}{\left\Vert \ket{\psi}\right\Vert_2}$:
\begin{equation}
    e^{iHt}=\mathbb{I} + iHt - \frac{H^2 t^2}{2}  - \frac{iH^3t^3}{6}+ \mathcal{O}(\tau^4).
\end{equation}
Substituting this relation into the trace argument of \eqref{eq:modifiedacc}, 
we find that
\begin{multline}
    \operatorname{Tr}[U^\dag e^{iHt} U e^{-iHt}] 
     = d + t^2(\operatorname{Tr}[H U^\dag H U] - \operatorname{Tr}[H^2]) \\+ \frac{it^3}{2}(\operatorname{Tr}[U^\dag H^2 UH] - \operatorname{Tr}[U^\dag H UH^2]) + \mathcal{O}(\tau^4),
\end{multline}
where the equality is obtained using the linearity and cyclicity properties of the trace. After summing over all group elements, as in \eqref{eq:modifiedacc}, and using the group property (that $g\in G$ implies $g^{-1} \in G$), we find that $\frac{1}{|G|} \sum _{g \in G}(\operatorname{Tr}[U^\dag(g) H^2 U(g)H] - \operatorname{Tr}[U^\dag(g) H U(g)H^2]) = 0$, so  that the third order term of \eqref{eq:modifiedacc} vanishes. We can simplify the second order term of \eqref{eq:modifiedacc} by using 
\begin{equation}
    \frac{1}{2}\operatorname{Tr}\!\left[|[U,H]|^2\right] =-\operatorname{Tr}[H U^\dag H U] + \operatorname{Tr}[H^2],
\end{equation}
where $|X|^2 \coloneqq X^\dag X$ implies that
\begin{equation}
    |[U,H]|^2 = H^2 - H U^\dag H U - U^\dag H U H + U^\dag H^2 U.
\end{equation}

Putting these equations together, we can  rewrite the acceptance probability of our first quantum algorithm elegantly as
\begin{equation}
    P_{\text{acc}} = 1 - \frac{t^2}{2 d |G|}\sum_{g \in G} \Big\|[U(g),H]\Big\|^2_2 + \mathcal{O}(\tau^4),
    \label{eq:accept-prob-comm}
\end{equation}
where $\left\| A \right\|_2 \coloneqq \sqrt{\operatorname{Tr}[|A|^2]}$ is the Hilbert--Schmidt norm.
Thus, to the first non-vanishing order of time $t$, the acceptance probability is equal to one if and only if
\begin{equation}
    [U(g),H] = 0, \quad \forall g \in G.
\end{equation}
This is exactly the familiar expression for symmetry. Furthermore, the expression in \eqref{eq:accept-prob-comm} clarifies that the normalized commutator norm $\frac{1}{ d |G|}\sum_{g \in G} \big\|[U(g),H]\big\|^2_2$ can be estimated efficiently by employing our algorithm. From \eqref{eq:accept-prob-comm}, we can see that the normalized commutator norm is small---equivalently, the Hamiltonian $H$ is approximately symmetric---if and only if the acceptance probability is close to one. (See \cite[Sections~III and V]{Alexander2022} or \cite{marvian2012symmetry} for further discussions on asymmetry fluctuations.)
Finally, as we show in Appendix~\ref{app:exact-expansion}
, the acceptance probability has an exact expansion as follows, such that all odd powers in $t$ vanish and the even powers are scaled by normalized nested commutator norms, quantifying higher orders of symmetry:
\begin{equation}
    P_{\text{acc}} =\sum_{n=0}^{\infty}\frac{\left(  -1\right)
^{n}t^{2n}}{ \left(  2n!\right)  }\left(\frac{1}{d\left\vert G\right\vert}\sum_{g\in
G}\Big\Vert \left[  \left(  H\right)  ^{n},U(g)\right]  \Big\Vert _{2}^{2}\right)
\label{eq:acc_prob_beauty}
\end{equation}
where the nested commutator is defined as
\begin{equation}
[(X)^n,Y]  \coloneqq \underbrace{[X,\dotsb[X,[X}_{n \text { times }}, Y]] \dotsb],\quad
\quad [(X)^0,Y]  \coloneqq Y.
\label{eq:def-nested-comm}
\end{equation}
Note that the expansion in \eqref{eq:acc_prob_beauty} is valid for all $t\in\mathbb{R}$.
We also provide an alternative formula for $P_{\text{acc}}$ in Appendix~\ref{app:exact-expansion}.

\textit{Variational Quantum Algorithm for Symmetry Testing}---Rather than feeding in the maximally-mixed state to the input of the circuit in Figure~\ref{fig:altcircuit}, we can instead feed in an arbitrary input state $\ket{\psi}$. As shown in Appendix~\ref{app:variational-sym-test}
, the acceptance probability when doing so is equal to
\begin{equation}
    \left\Vert \mathcal{T}_G(e^{-iHt})
|\psi\rangle\right\Vert _{2}^{2} =
1-t^{2}\left\langle \mathcal{T}_{G}(H^{2})-\left(  \mathcal{T}%
_{G}(H)\right)  ^{2}\right\rangle _{\psi}+O(\tau^{3}),
\label{eq:accept-prob-fixed-state}
\end{equation}
where $
\mathcal{T}_{G}(X)\coloneqq \frac{1}{|G|}\sum_{g\in G}U(g)XU^{\dag}(g)$.
Note that the bracketed term is non-negative as a consequence of the Kadison--Schwarz inequality \cite[Theorem~2.3.2]{bhatia07positivedefinitematrices}.
If we had the ability to prepare arbitrary quantum states (modeled in \cite{VW15}), we could optimize this acceptance probability over all states, resulting in the following value:
\begin{align}
    \left\Vert \mathcal{T}_G(e^{-iHt})\right\Vert _{\infty}^{2}   & \geq
1-\frac{2}{\left\vert G\right\vert }\sum_{g\in G}\left\Vert \left[
U(g),e^{-iHt}\right]  \right\Vert _{\infty} 
\label{eq:accept-prob-optimized} \\
& \geq 
1-\frac{2t}{\left\vert G\right\vert }\sum_{g\in G}\left\Vert \left[
U(g),H\right]  \right\Vert _{\infty}-4\tau^{2}
\label{eq:lower-bnd-est-small-t}.
\end{align}
These inequalities are proven in Appendix~\ref{app:variational-sym-test}, 
and the second holds under the assumption that $ \tau < 1$. This
demonstrates that the acceptance probability $\left\Vert \mathcal{T}_G(e^{-iHt})\right\Vert _{\infty}^{2}$ can be bounded from below in terms of a familiar expression of Hamiltonian symmetry. Thus, if the commutator norm $\frac{1}{\left\vert G\right\vert }\sum_{g\in G}\left\Vert \left[
U(g),H\right]  \right\Vert _{\infty}$ is small, as is the case when the Hamiltonian is approximately symmetric, then the acceptance probability of this algorithm is close to one. In Appendix~\ref{app:variational-sym-test}, 
we also prove that the acceptance probability satisfies
\begin{equation}
\left\Vert \mathcal{T}_G(e^{-iHt})\right\Vert _{\infty}^{2} \geq
    \left(  1-\sum_{n=1}^{\infty}\frac{t^{n}}{n!}\frac{1}{\left\vert
G\right\vert }\sum_{g\in G}\Big\Vert \left[  \left(  H\right)  ^{n}%
,U(g)\right]  \Big\Vert _{\infty}\right)  ^{2}.
\label{eq:qma-nested-comm-bnd}
\end{equation}

Since it is physically impossible to optimize over all input states, we can instead employ a variational ansatz to do so, in order to arrive at a lower bound estimate of the acceptance probability on the left-hand side of \eqref{eq:accept-prob-optimized}. These methods have been vigorously pursued in recent years in the quantum computing literature \cite{CABBEFMMYCC20,bharti2021noisy}, and they can be combined with our approach here. In short, the acceptance probability in \eqref{eq:accept-prob-fixed-state} is a reward function that can be estimated by means of the circuit in Figure~\ref{fig:altcircuit} and a parameterized circuit that prepares the state $\ket{\psi}$. Then one can employ gradient ascent on a classical computer to modify the parameters used to prepare the state $\ket{\psi}$. After many iterations, these algorithms typically converge to a value, which in our case provides a lower bound estimate of the acceptance probability on the left-hand side of~\eqref{eq:accept-prob-optimized}. In practice, it might be difficult in experiments to optimize over all pure states, and one could instead consider a variational product state ansatz, as in \cite{GLXXGLSPXZWZF21}. 

\begin{figure}[t!]
\begin{center}
\includegraphics[width=3.4in]{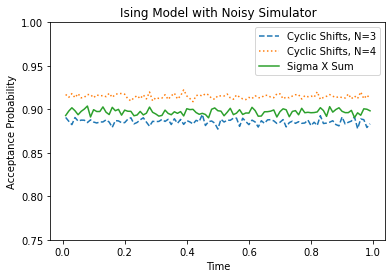}
\end{center}
\caption{Results of symmetry tests for the transverse Ising model for $N=3$ and $N=4$, using IBM Quantum's noisy simulator. The symmetries in question are given by acting simultaneously on all systems by either the cyclic group of order $N$ or a conjugation by~$(\sigma^X)^{\otimes N}$.}
\label{fig:ising}
\end{figure}

\textit{\label{sec:level5}Examples}---To exhibit our algorithm, we consider the dynamics given by the transverse Ising model with a cyclic boundary condition. This Hamiltonian is given as $
    H_{\text{TIM}} \coloneqq \sigma^Z_N \otimes \sigma^Z_1 + \sum_{i=1}^{N-1} \sigma^Z_i \otimes \sigma^Z_{i+1} +  \sum_{i=1}^N \sigma^X_i$.
This Hamiltonian is permutationally invariant, so that $[H_{\text{TIM}},W^{\pi}] = 0$ for all $\pi \in S_N$, where $W^{\pi}$ is a unitary representation of the permutation $\pi \in S_N$, with $S_N$ denoting the symmetric group of $N$ elements. It also obeys the symmetry $[H_{\text{TIM}},\sigma^X_1 \otimes \cdots \otimes \sigma^X_N] = 0$. We thus can use our algorithm to test these symmetries, and we do so in Figure~\ref{fig:ising} for $N=3$ and $N=4$. (Rather than test all permutations, we indicate here that we test for invariance under a cyclic shift.) We find that each respective symmetry test passes with reasonable probability, with deviation from one due to noise added to the simulation. In Appendix~\ref{app:extra}, 
we implement symmetry tests for two other examples--- the weakly $J$-coupled NMR Hamiltonian and the Heisenberg XY model. We note here that all computer codes used to generate the examples in the main text and the supplementary material are available online at github\footnote{\url{https://github.com/mlabo15/Hamiltonian-Symmetry}} and as arXiv ancillary files.

\textit{Conclusion}---In this work, we have specified algorithms to test a Hamiltonian for symmetry with respect to a group.
Our first test is efficiently realizable given similarly efficient Hamiltonian simulations, and our second test employs a variational approach. 
These algorithms are useful tools that we suspect should be of  interest throughout many realms of physics.

\begin{acknowledgments}
MLL acknowledges support from the DOD Smart Scholarship program, and MMW from NSF Grant No.~1907615. We are grateful to Bryan Gard, Zo\"e Holmes, and Soorya Rethinasamy for feedback on our manuscript. We acknowledge use of the IBM~Q for this work. The views expressed are those of the authors and do not reflect the official policy or position of IBM or the IBM~Q team.
\end{acknowledgments}

\bibliography{main}
\bibliographystyle{unsrt}

\appendix

\section{Covariance of a Quantum Channel in Detail}
\label{app:covariance}
 Let $G$ be a group with projective unitary representations $\{U(g)_A\}_{g \in G}$ and $\{V(g)_B\}_{g \in G}$ on the $A$ and $B$ subsystems respectively. Then the channel $\mathcal{N}_{A\rightarrow B}$ is covariant if the following $G$-covariance symmetry condition holds
\begin{equation}
\mathcal{N}_{A\rightarrow B}\circ\mathcal{U}_{A}(g)=\mathcal{V}_{B}(g) \circ \mathcal{N}_{A\rightarrow B} \qquad \forall g\in G,
\label{eq:ch-cov-def}
\end{equation}
where the unitary channels $\mathcal{U}_{A}(g)$ and $\mathcal{V}_{B}(g)$ are
respectively defined from $U_{A}(g)$ and $V_{B}(g)$ as%
\begin{align}\label{eq:uandv}
\mathcal{U}_{A}(g)(\omega_{A}) &  \coloneqq  U_{A}(g)\omega_{A}U_{A}(g)^{\dag
},\\
\mathcal{V}_{B}(g)(\tau_{B}) &  \coloneqq  V_{B}(g)\tau_{B}V_{B}(g)^{\dag}.
\end{align}

Furthermore, a channel is covariant in the sense above if and only if its Choi state is invariant in the following sense \cite[Eq.~(59)]{CDP09}:
\begin{equation} \label{eq:choicondition}
\Phi_{RB}^{\mathcal{N}}=(\overline{\mathcal{U}}_{R}(g)\otimes\mathcal{V}
_{B}(g))(\Phi_{RB}^{\mathcal{N}})\quad\forall g\in G,
\end{equation}
where
\begin{equation}
\overline{\mathcal{U}}_{R}(g)(\omega_{R})\coloneqq  \overline{U}_{R}
(g)\omega_{R}U_{R}(g)^{T},
\end{equation}
where we have defined the Choi state, $\Phi_{RB}^{\mathcal{N}}$, of the channel $\mathcal{N}_{A\rightarrow B}$ as 
\begin{align}
\Phi_{RB}^{\mathcal{N}}  & \coloneqq  \mathcal{N}_{A\rightarrow B}(\Phi_{RA}),\\
\Phi_{RA}  & \coloneqq  \frac{1}{\left\vert A\right\vert }\sum_{i,j}|i\rangle\!\langle
j|_{R}\otimes|i\rangle\!\langle j|_{A}.
\end{align}
We can note a different condition by specifying a projector over the space of states symmetric with respect to group $G$ \cite{harrow2013church}. Denote the projector of this group as
\begin{equation}\label{eq:projector_app}
    \Pi^G =\frac{1}{|G|} \sum_{g \in G} \overline{U}_R(g) \otimes U_B (g) .
\end{equation}
Then the Choi state is symmetric with respect to $G$ if
\begin{equation}\label{eq:bosechoi}
    \Phi_{RB}^{\mathcal{N}} = \Pi^G \Phi_{RB}^{\mathcal{N}}.
\end{equation}
The above definition is a stronger condition of symmetry; thus, if the Choi state obeys \eqref{eq:bosechoi} then \eqref{eq:choicondition} follows.

If a channel describing Hamiltonian dynamics exhibits $G$-covariant symmetry, then the underlying Hamiltonian is symmetric with respect to $G$. This can be seen by manipulating \eqref{eq:ch-cov-def} using the definition of the Choi state and the covariance condition.

\section{Approximate Symmetry and Continuity in the Acceptance Probability}

\label{app:approximate}

In this appendix, we consider approximate versions of the statement
\begin{equation}
\operatorname{Tr}[\Pi^{G}\Phi_{RB}^{t}]=1\quad\Leftrightarrow\quad \Phi_{RB}^{t}
=\Pi^{G}\Phi_{RB}^{t}\Pi^{G},
\label{eq:app-sym-conditions}
\end{equation}
in which we investigate  cases in which the Choi state is only approximately symmetric.

Let us begin with the first version and set
\begin{equation}
    \epsilon \coloneqq \left\Vert \Phi_{RB}^{t} - \Pi^{G}\Phi_{RB}^{t}\Pi^{G} \right\Vert_1 \, .
    \label{eq:approx-sym-def}
\end{equation}
Here we employ the trace distance as a standard metric between states or subnormalized states.
%
From this point, we use the reverse triangle inequality to conclude that
\begin{equation}
    \left\Vert \Phi_{RB}^{t} - \Pi^{G}\Phi_{RB}^{t}\Pi^{G} \right\Vert_1 \geq \left\Vert \Phi_{RB}^{t} \right\Vert_1 - \left\Vert \Pi^{G}\Phi_{RB}^{t}\Pi^{G} \right\Vert_1,
    \label{eq:rev-tri-app}
\end{equation}
where the first term on the right-hand side is equal to one for every quantum state, and the left-hand side is equal to $\epsilon$. Meanwhile, recall that our acceptance probability is given by
\begin{equation}
    P_{\textrm{acc}}=\operatorname{Tr}[ \Pi^{G}\Phi_{RB}^{t}]=\operatorname{Tr}[ \Pi^{G}\Phi_{RB}^{t}\Pi^{G}] = \left \Vert \Pi^{G}\Phi_{RB}^{t}\Pi^{G}  \right \Vert_1,
\end{equation}
where the second equality follows from cyclicity of trace and the last equality follows because $\Pi^{G}\Phi_{RB}^{t}\Pi^{G}$ is positive semi-definite.
This is the final term on the right-hand side of \eqref{eq:rev-tri-app}. Thus we conclude that
\begin{equation}
    P_{\textrm{acc}} \geq  1 - \epsilon\, .
\end{equation}

The consequence of this result is that whenever the state is approximately symmetric (in the sense that $\epsilon \approx 0$ in \eqref{eq:approx-sym-def}), the acceptance probability is near to one. Thus, our acceptance probability demonstrates a continuity property.

Next, we will show that the reverse direction is also true. This relationship can be demonstrated via the gentle operator lemma \cite{Davies1969,itit1999winter,ON07}. Let $\Phi$ be a density operator  and $\Lambda$ a measurement operator satisfying $0 \leq \Lambda \leq I$.  If $\operatorname{Tr}[\Lambda \Phi] \geq 1 - \epsilon$ for $\epsilon \in [0,1]$, then the following inequality holds 
\begin{equation}
    \left \Vert \Phi - \sqrt{\Lambda}\Phi \sqrt{\Lambda} \right \Vert_1 \leq 2 \sqrt{\epsilon}\,.
\end{equation}

In our case, set $\Lambda = \Pi ^G$. (Note that $\Pi^G = \sqrt{\Pi^G}$ for a projector.) Then suppose our acceptance probability is close to one, as in
\begin{equation}
    P_{\textrm{acc}}=\operatorname{Tr}[\Pi^G \Phi^t_{RB}] \geq 1 - \epsilon\, ,
\end{equation}
for some $\epsilon \in [0,1]$. Then the gentle operator lemma implies that the state is approximately symmetric:
\begin{equation}
   \left \Vert \Phi^t_{RB} - \Pi^G \Phi^t_{RB} \Pi^G \right \Vert_1 \leq 2  \sqrt{\epsilon} \, .
\end{equation}

Another approximate version of the statement in \eqref{eq:app-sym-conditions} is as follows. Let us now set
\begin{equation}
    \epsilon \coloneqq \frac{1}{2} \left \Vert \Phi^t_{RB} - \rho^t_{RB} \right \Vert_1 \in [0,1],
\end{equation}
where $\rho^t_{RB}$ is defined as the post-measurement state after the algorithm accepts:
\begin{equation}
\rho^t_{RB} \coloneqq \frac{\Pi^G \Phi^t_{RB}\Pi^G }{\operatorname{Tr}[\Pi^G \Phi^t_{RB}]}.
\end{equation}
By the well known variational characterization of the normalized trace distance of states $\omega$ and $\tau$ as $\frac{1}{2} \left \Vert  \omega - \tau \right \Vert_1 = \max_{\Lambda : 0\leq \Lambda \leq I} \operatorname{Tr}[\Lambda(\tau -\omega)]$, it follows that
\begin{align}
    \epsilon & =  \frac{1}{2} \left \Vert \Phi^t_{RB} - \rho^t_{RB} \right \Vert_1 \\
    & = \max_{\Lambda_{RB} : 0\leq \Lambda_{RB} \leq I_{RB}} \operatorname{Tr}[\Lambda_{RB}(\rho^t_{RB} -\Phi^t_{RB})] \\
    & \geq \operatorname{Tr}[\Pi^G(\rho^t_{RB} -\Phi^t_{RB})]\\
    & = 1 - \operatorname{Tr}[\Pi^G\Phi^t_{RB}]\\
    & = 1-P_{\text{acc}}.
\end{align}
Thus we conclude that
\begin{equation}
    P_{\text{acc}} \geq 1- \epsilon.
\end{equation}
For the reverse direction, recall the gentle measurement lemma \cite[Lemma~9.4.1]{Wbook17}: Let $\Phi$ be a density operator  and $\Lambda$ a measurement operator satisfying $0 \leq \Lambda \leq I$.  If $\operatorname{Tr}[\Lambda \Phi] \geq 1 - \epsilon$ for $\epsilon \in [0,1]$, then the following inequality holds 
\begin{equation}
    \frac{1}{2} \left \Vert \Phi - \frac{\sqrt{\Lambda}\Phi \sqrt{\Lambda}}{\operatorname{Tr}[\Lambda \Phi]} \right \Vert_1 \leq  \sqrt{\epsilon}\,.
\end{equation}
Applying this, let us suppose that the acceptance probability is close to one, as in
\begin{equation}
    P_{\textrm{acc}}=\operatorname{Tr}[\Pi^G \Phi^t_{RB}] \geq 1 - \epsilon\, ,
\end{equation}
for some $\epsilon \in [0,1]$. Then the gentle measurement lemma implies that the state is approximately symmetric, in the following sense:
\begin{equation}
   \frac{1}{2} \left \Vert \Phi^t_{RB} - \rho^t_{RB} \right \Vert_1 \leq   \sqrt{\epsilon} \, .
\end{equation}

\section{Acceptance Probability of the First (Efficient) Hamiltonian Symmetry Test}

\label{app:accept}

To see that the acceptance probability of the first Hamiltonian symmetry test in Figure~\ref{fig:originalcircuit}
is given by $\operatorname{Tr}[\Pi^{G}\Phi_{RB}^{t}]$, consider that the state
just before the measurement is as follows:
\begin{equation}
\frac{1}{\sqrt{\left\vert G\right\vert }}\sum_{g\in G}|g\rangle_{C}\left(
\overline{U}_{R}(g)\otimes U_{B}(g)\right)  |\Phi^{t}\rangle_{RB}.
\end{equation}
Then the acceptance probability is given by
\begin{align}
& \left\Vert
\begin{array}
[c]{c}
\left(  \langle+|_{C}\otimes \mathbb{I}_{RB}\right)  \times\\
\left(  \frac{1}{\sqrt{\left\vert G\right\vert }}\sum_{g\in G}|g\rangle
_{C}\left(  \overline{U}_{R}(g)\otimes U_{B}(g)\right)  |\Phi^{t}\rangle
_{RB}\right)
\end{array}
\right\Vert _{2}^{2}\nonumber\\
& =\left\Vert
\begin{array}
[c]{c}
\left(  \frac{1}{\sqrt{\left\vert G\right\vert }}\sum_{g^{\prime}\in G}\langle
g^{\prime}|_{C}\otimes \mathbb{I}_{RB}\right)  \times\\
\left(  \frac{1}{\sqrt{\left\vert G\right\vert }}\sum_{g\in G}|g\rangle
_{C}\left(  \overline{U}_{R}(g)\otimes U_{B}(g)\right)  |\Phi^{t}\rangle
_{RB}\right)
\end{array}
\right\Vert _{2}^{2}\\
& =\left\Vert \frac{1}{\left\vert G\right\vert }\sum_{g^{\prime},g\in
G}\langle g^{\prime}|g\rangle_{C}\left(  \overline{U}_{R}(g)\otimes
U_{B}(g)\right)  |\Phi^{t}\rangle_{RB}\right\Vert _{2}^{2}\\
& =\left\Vert \frac{1}{\left\vert G\right\vert }\sum_{g\in G}\left(
\overline{U}_{R}(g)\otimes U_{B}(g)\right)  |\Phi^{t}\rangle_{RB}\right\Vert
_{2}^{2}\\
& =\left\Vert \Pi^{G}|\Phi^{t}\rangle_{RB}\right\Vert _{2}^{2}\\
& =\operatorname{Tr}[\Pi^{G}\Phi_{RB}^{t}].
\end{align}

Now we show that 
\begin{equation}\label{eq:start}
    P_{\text{acc}} = \operatorname{Tr}[\Pi^G \Phi_{RB}^{t}]
\end{equation}
is equal to the following expression:
\begin{equation}
    P_{\text{acc}}=\frac{1}{d |G|}\sum_{g \in G}\operatorname{Tr}[U^\dag (g) e^{i H t} U(g) e^{-i H t}]\, .
\end{equation}
To see this, we begin with equation \eqref{eq:start} and note that, using the cyclicity of the trace, it can be rewritten as
\begin{equation}
    \operatorname{Tr}[(\mathbb{I}_R \otimes e^{iHt}) \Pi^G (\mathbb{I}_R \otimes e^{-iHt}) \Phi_{RA}].
\end{equation}
We can now substitute the definition of the projector given in \eqref{eq:projector_app}, giving
\begin{multline}
    \frac{1}{|G|}\operatorname{Tr}[(\mathbb{I}_R \otimes e^{iHt}) \sum_{g \in G} \overline{U}_R(g) \otimes U_A(g) (\mathbb{I}_R \otimes e^{-iHt}) \Phi_{RA}], \\
    = \frac{1}{|G|}\sum_{g \in G} \operatorname{Tr}[(\mathbb{I}_R \otimes e^{iHt}) \overline{U}_R(g) \otimes U_A(g) (\mathbb{I}_R \otimes e^{-iHt}) \Phi_{RA}],
\end{multline}
where the second equality follows from the linearity of the trace. 

We now want to employ the transpose trick:
\begin{equation}
    \mathbb{I}_R \otimes U_A \ket{\Phi}_{RA} = U^{T}_R \otimes \mathbb{I}_A \ket{\Phi}_{RA},
\end{equation}
where $T$ denotes the transpose. The description of this action can be easily interpreted through the language of tensor networks \cite{biamonte2017tensor}. Using this relation, we can rewrite the above as
\begin{equation}
    \frac{1}{|G|}\sum_{g \in G} \operatorname{Tr}[U^\dag_A(g)(\mathbb{I}_R \otimes e^{iHt}) U_A(g) (\mathbb{I}_R \otimes e^{-iHt}) \Phi_{RA}].
\end{equation}

We can now evaluate the trace as a composition of partial traces ($\operatorname{Tr} [\cdot] =\operatorname{Tr}_{RA}[\cdot] = \operatorname{Tr}_A [\operatorname{Tr}_R[\cdot]]$). Computing the trace on $R$ first, we obtain
\begin{equation}
    \frac{1}{d |G|}\sum_{g \in G} \operatorname{Tr}[U^\dag_A(g) e^{iHt} U_A(g) e^{-iHt} ],
\end{equation}
which is exactly  \eqref{eq:modifiedacc}. 

\section{DQC1-completeness of acceptance probability}

\label{sec:DQC1-completeness}

In this appendix, we prove that estimating the acceptance probability in \eqref{eq:modifiedacc} to within additive error is a DQC1-complete  problem. 

Before proving this result, let us first review some basics of the DQC1 complexity class \cite{KL98,SJ08}. The basic model involves preparing one qubit in a pure state $\ket{0}\!\bra{0}$ and all other qubits (suppose there are $n$ of them) in the maximally mixed state $I/d$, where $d\coloneqq 2^n$. Then a quantum circuit, which generates a unitary $U$, is performed on all of the qubits, and the first qubit is measured in the computational basis $\{|0\rangle\!\langle 0|,|1\rangle\!\langle 1|\}$. The algorithm accepts if the outcome $|1\rangle\!\langle 1|$ occurs. The problem of estimating the acceptance probability $\operatorname{Tr}\!\left[(|1\rangle\!\langle 1|\otimes I)U(|0\rangle\!\langle 0|\otimes I/d)U^\dag\right]$ to within additive error is a DQC1-complete problem, by definition.

One of the insights of \cite[Section~1]{SJ08} is that the problem of estimating $\operatorname{Re}[\operatorname{Tr}[U]]/d$ to within additive error, where $U$ is the unitary realized by a quantum circuit acting on $n$ qubits and consisting of polynomially many gates, is a DQC1-complete problem. This means that the problem can be solved within the computational model mentioned above, and it is also just as hard as every other problem that can be solved in the model. Thus, in this sense, this problem of normalized trace estimation characterizes the class DQC1. Another key observation of \cite[Section~1]{SJ08} is that the complexity class DQC1 does not change if there are a constant or even logarithmic number of pure qubits, where here we mean logarithmic in $n$.

Before establishing the main result of this appendix, we first prove that estimating $\operatorname{Re}[\operatorname{Tr}[U^{2}]]/d$, where $U$ is the unitary generated by a quantum circuit, is a
DQC1-complete problem. Here we make use of the aforementioned fact that estimating
$\operatorname{Re}[\operatorname{Tr}[U]]/d$ is a DQC1-complete 
problem. First, to show that estimating $\operatorname{Re}[\operatorname{Tr}%
[U^{2}]]/d$ is in DQC1, we can use the usual construction (see \cite[Section~1]{SJ08}) with $U$ substituted
by $U^{2}$. Indeed, we prepare a control qubit in the $|+\rangle$ state and
all other qubits in the maximally mixed state, act with controlled-$U^{2}$ (easily realized as two applications of controlled-$U$),
and then measure in the Hadamard basis. Assigning the values $+1$ and $-1$ to the measurement outcomes, the expected value of the measurement outcomes is equal to $\operatorname{Re}[\operatorname{Tr}[U^{2}]]/d$. This implies that the problem is in DQC1.
To show hardness, suppose that we have a way of estimating $\operatorname{Re}%
[\operatorname{Tr}[U^{2}]]/d$ for every $U$, where $U$ is the unitary
generated by a quantum circuit. Then we show that it possible to use such an algorithm to estimate
$\operatorname{Re}[\operatorname{Tr}[U]]/d$. The key idea behind the reduction
is the following unitary%
\begin{equation}
V:=%
\begin{bmatrix}
0 & I\\
U & 0
\end{bmatrix}
=|0\rangle\!\langle1|\otimes I+|1\rangle\!\langle0|\otimes U,
\end{equation}
which can be realized in terms of a quantum circuit as a $\sigma_{X}$ acting on a control qubit, followed by
a controlled-$U$, i.e.,%
\begin{equation}
\left(  |0\rangle\!\langle0|\otimes I+|1\rangle\!\langle1|\otimes U\right)
\left(  \sigma_{X}\otimes I\right)  .
\end{equation}
We then observe that%
\begin{align}
\operatorname{Tr}\left[  V^{2}\right]   &  =\operatorname{Tr}\left[
\begin{bmatrix}
0 & I\\
U & 0
\end{bmatrix}%
\begin{bmatrix}
0 & I\\
U & 0
\end{bmatrix}
\right]  \\
&  =\operatorname{Tr}\left[
\begin{bmatrix}
U & 0\\
0 & U
\end{bmatrix}
\right]  \\
&  =2\operatorname{Tr}[U].
\end{align}
Thus, by using the method to estimate $\operatorname{Re}[\operatorname{Tr}%
[V^{2}]]/d$, we estimate $\operatorname{Re}[\operatorname{Tr}[U]]/d$ up to a
constant factor of $2$. This completes the proof that estimating
$\operatorname{Re}[\operatorname{Tr}[U^{2}]]/d$ is a DQC1-complete problem.

Now let us turn to the main goal of this appendix: proving that estimating the acceptance probability in \eqref{eq:modifiedacc} to within additive error is a DQC1-complete problem. Let $H$ be a local Hamiltonian, and let $\left\{
U(g)\right\}  _{g\in G}$ be a unitary representation of a group $G$, such that
each $U(g)$ can be realized by a quantum circuit. Furthermore, suppose that the size $|G|$ of the group~$G$ is no larger than linear in the number of qubits on which each circuit for $U(g)$ acts (so that $\log|G|$ is logarithmic in the number of qubits).  The size of the
computational problem is in terms of the classical description of the
Hamiltonian $H$ and the circuit descriptions of each unitary $U(g)$. The task
of estimating the value
\begin{equation}
\frac{1}{d\left\vert G\right\vert }\sum_{g\in G}\operatorname{Tr}[U^{\dag
}(g)e^{iHt}U(g)e^{-iHt}]\label{eq:acceptance-prob-alg}%
\end{equation}
is a DQC1-complete problem. To show that this is in DQC1, we use the quantum
algorithm presented in our paper, as well as the observation from \cite[Section~1]{SJ08}, that the class DQC1 does not change if there are a logarithmic number of pure qubits (which is the case under our description of the problem stated above). To show hardness, let $U$ be a unitary
realized by an arbitrary quantum circuit, acting on a $d'$-dimensional Hilbert space, and we use the fact that estimating
$\operatorname{Re}[\operatorname{Tr}[U^{2}]]/d'$ is a DQC1-complete problem (as
proven above). We show that an algorithm for estimating the value in
\eqref{eq:acceptance-prob-alg} can estimate the value $\operatorname{Re}%
[\operatorname{Tr}[U^{2}]]/d'$. To this end, we let the group $G$ be $Z_{2}$
with representation $\left\{  I,V\right\}  $, where%
\begin{equation}
V=|0\rangle\!\langle1|\otimes U+|1\rangle\!\langle0|\otimes U^{\dag}.
\label{eq:V_def}
\end{equation}
Note that $V$ acts on a $d = 2d'$-dimensional Hilbert space.
(A quick calculation indicates that $V^2=I$, so that indeed $\{I,V\}$ is a representation of $Z_2$.)
A circuit for realizing $V$ can be efficiently generated from the circuit for realizing
$U$. Indeed, we can construct a $0$-controlled-$U$ from~$U$:%
\begin{equation}
|0\rangle\!\langle0|\otimes U+|1\rangle\!\langle1|\otimes I,
\end{equation}
and a $1$-controlled-$U^{\dag}$ from $U$, by running $U$ backwards with each
circuit gate controlled on $1$, leading to
\begin{equation}
|0\rangle\!\langle0|\otimes I+|1\rangle\!\langle1|\otimes U^{\dag}.
\end{equation}
The overall circuit consists of $X\otimes I$, and then the above controlled
gates, so that%
\begin{align}
&  \left(  |0\rangle\!\langle0|\otimes U+|1\rangle\!\langle1|\otimes I\right)
\left(  |0\rangle\!\langle0|\otimes I+|1\rangle\!\langle1|\otimes U^{\dag
}\right)  \left(  X\otimes I\right)  \nonumber\\
&  =\left(  |0\rangle\!\langle0|\otimes U+|1\rangle\!\langle1|\otimes U^{\dag
}\right)  \left(  X\otimes I\right)  \\
&  =|0\rangle\!\langle1|\otimes U+|1\rangle\!\langle0|\otimes U^{\dag}\\
&  =V.
\end{align}
We take the Hamiltonian to be one that realizes $H_{2}\otimes I$ via Hamiltonian evolution, where
$H_{2}$ is a $2\times2$ Hadamard gate. Indeed such a Hamiltonian is local, acting non-trivially on only one qubit. Then we have that $\left\vert
G\right\vert =2$, its unitary representation is $\left\{  I,V\right\}  $, and
$e^{-iHt}=H_{2}\otimes I$. Plugging into \eqref{eq:acceptance-prob-alg}, we
find that%
\begin{multline}
\frac{1}{d\left\vert G\right\vert }\sum_{g\in G}\operatorname{Tr}[U^{\dag
}(g)e^{iHt}U(g)e^{-iHt}]\label{eq:step-0}\\
=\frac{1}{4d'}\operatorname{Tr}\left[  I\left(  H_{2}\otimes I\right)  I\left(
H_{2}\otimes I\right)  \right]  \\
+\frac{1}{4d'}\operatorname{Tr}\left[  V\left(  H_{2}\otimes I\right)  V\left(
H_{2}\otimes I\right)  \right]  .
\end{multline}
Consider that%
\begin{equation}
\operatorname{Tr}\left[  I\left(  H_{2}\otimes I\right)  I\left(  H_{2}\otimes
I\right)  \right]  =\operatorname{Tr}[I_2 \otimes I_{d'}] = 2d',\label{eq:step-1}%
\end{equation}
and\begin{widetext}%
\begin{align}
&  \operatorname{Tr}\left[  V\left(  H_{2}\otimes I\right)  V\left(
H_{2}\otimes I\right)  \right]  \nonumber\\
&  =\operatorname{Tr}\left[  \left(  |0\rangle\!\langle1|\otimes
U+|1\rangle\langle0|\otimes U^{\dag}\right)  \left(  H_{2}\otimes I\right)
\left(  |0\rangle\!\langle1|\otimes U+|1\rangle\!\langle0|\otimes U^{\dag
}\right)  \left(  H_{2}\otimes I\right)  \right]  \label{eq:step-2}\\
&  =\operatorname{Tr}\left[  \left(  |0\rangle\!\langle1|\otimes
U+|1\rangle\langle0|\otimes U^{\dag}\right)  \left(  |+\rangle\!\langle
-|\otimes U+|-\rangle\!\langle+|\otimes U^{\dag}\right)  \right]  \\
&  =\operatorname{Tr}\left[  |0\rangle\!\langle1|+\rangle\!\langle-|\otimes
U^{2}+|1\rangle\!\langle0|+\rangle\!\langle-|\otimes I_{d'}+|0\rangle
\!\langle1|-\rangle\!\langle+|\otimes I_{d'}+|1\rangle\!\langle0|-\rangle
\!\langle+|\otimes\left(  U^{\dag}\right)  ^{2}\right]  \\
&  =\langle1|+\rangle\!\langle-|0\rangle\operatorname{Tr}\left[  U^{2}\right]
+\langle0|+\rangle\!\langle-|1\rangle\operatorname{Tr}\left[  I_{d'}\right]
+\langle1|-\rangle\!\langle+|0\rangle\operatorname{Tr}\left[  I_{d'}\right]
+\langle0|-\rangle\!\langle+|1\rangle\operatorname{Tr}\left[  \left(  U^{\dag
}\right)  ^{2}\right]  \\
&  =\frac{1}{2}\left(  \operatorname{Tr}\left[  U^{2}\right]
-\operatorname{Tr}\left[  I_{d'}\right]  -\operatorname{Tr}\left[  I_{d'}\right]
+\operatorname{Tr}\left[  \left(  U^{\dag}\right)  ^{2}\right]  \right)\\
& =\frac{1}{2}\left(  \operatorname{Tr}\left[  U^{2}\right]
-2d'
+\operatorname{Tr}\left[  \left(  U^{\dag}\right)  ^{2}\right]  \right) \\
& =  \operatorname{Re}\left[  \operatorname{Tr}%
\left[  U^{2}\right]\right] - d'
.\label{eq:step-2b}%
\end{align}
\end{widetext}
Putting together \eqref{eq:step-0}, \eqref{eq:step-1}, and
\eqref{eq:step-2}--\eqref{eq:step-2b}, this implies for the above choices that%
\begin{multline}
\frac{1}{d\left\vert G\right\vert }\sum_{g\in G}\operatorname{Tr}[U^{\dag
}(g)e^{iHt}U(g)e^{-iHt}]\\
=\frac{1}{4} + \frac{\operatorname{Re}\left[  \operatorname{Tr}%
\left[  U^{2}\right]  \right]  }{4d'}.
\end{multline}
Thus, the acceptance probability of this algorithm provides an estimate of the
desired quantity, up to an additive factor of $1/4$ and a constant scaling factor of $1/2$, establishing
DQC1-hardness of our problem.

In our proof, we have placed the computational complexity of the original quantum circuit into the unitary representation of the group, via the unitary $V$ in \eqref{eq:V_def}, while having a rather trivial Hamiltonian. It is an interesting open problem to place the complexity of the original circuit into the Hamiltonian, rather than the group representation. It seems plausible that one could accomplish this via the well known Feynman circuit-to-Hamiltonian construction \cite{F85}. In more detail, the group could again be $Z_2$ but with the simple representation $\{I,\text{SWAP}\}$, whereas the Hamiltonian evolution would be chosen such that $U\otimes I=e^{-iHt}$, where $U$ is the original quantum circuit. Then one can verify that the acceptance probability in \eqref{eq:V_def} evaluates to
\begin{multline}
\frac{1}{4}\left \Vert (U_{A_1} + U_{A_2}) \ket{\Phi}_{A_1B_1}\ket{\Phi}_{A_2B_2}\right \Vert_2^2 = \frac{1}{2}(1 + \left|\operatorname{Tr}[U]/d\right|^2).
\end{multline}
Key technical questions here are the details of realizing the equality $U\otimes I=e^{-iHt}$ and also establishing that evaluating $\left|\operatorname{Tr}[U]\right|^2$ is a DQC1-complete problem. For the first question, it seems necessary to incorporate a clock register, as in \cite{F85}, while at the same time maintaining the structure of the computational task.
We leave this interesting question for future work. 

\section{Exact Expansion of the Acceptance Probability of the First (Efficient) Hamiltonian Symmetry Test}

\label{app:exact-expansion}

Here we first prove that the following equality holds:%
\begin{multline}
\frac{1}{d\left\vert G\right\vert }\sum_{g\in G}\operatorname{Tr}[U^{\dag
}(g)e^{iHt}U(g)e^{-iHt}]\\
=\frac{1}{d}\sum_{n=0}^{\infty}\frac{\left(  -1\right)  ^{n}}{\left(
2n\right)  !}t^{2n}f(n,k,H,G),
\end{multline}
where%
\begin{multline}
f(n,k,H,G)\coloneqq \\
\sum_{k=0}^{n}\binom{2n}{k}\left(  2-\delta_{k,n}\right)  \left(  -1\right)
^{k}\operatorname{Tr}[\mathcal{T}_{G}(H^{2n-k})H^{k}]
\end{multline}
and the group twirl $\mathcal{T}_{G}$ is defined as
\begin{equation}
\mathcal{T}_{G}(X)\coloneqq \frac{1}{|G|} \sum_{g\in G}U(g)XU^{\dag}(g).
\end{equation}
After that, we establish the expansion in \eqref{eq:acc_prob_beauty}.

Consider that%
\begin{align}
&  \frac{1}{\left\vert G\right\vert }\sum_{g}\operatorname{Tr}[U^{\dag
}(g)e^{iHt}U(g)e^{-iHt}]\nonumber\\
&  =\operatorname{Tr}[\mathcal{T}_{G}(e^{iHt})e^{-iHt}]\\
&  =\sum_{\ell=0}^{\infty}\frac{\left(  it\right)  ^{\ell}}{\ell
!}\operatorname{Tr}[\mathcal{T}_{G}(H^{\ell})e^{-iHt}]\\
&  =\sum_{\ell,m=0}^{\infty}\frac{\left(  it\right)  ^{\ell}\left(
-it\right)  ^{m}}{\ell!m!}\operatorname{Tr}[\mathcal{T}_{G}(H^{\ell})H^{m}]\\
&  =\sum_{\ell,m=0}^{\infty}\frac{\left(  it\right)  ^{\ell+m}\left(
-1\right)  ^{m}}{\ell!m!}\operatorname{Tr}[\mathcal{T}_{G}(H^{\ell})H^{m}]\\
&  =\sum_{n=0}^{\infty}\sum_{k=0}^{n}\frac{\left(  it\right)  ^{n}\left(
-1\right)  ^{k}}{n-k!k!}\operatorname{Tr}[\mathcal{T}_{G}(H^{n-k})H^{k}]\\
&  =\sum_{n=0}^{\infty}\left(  it\right)  ^{n}\sum_{k=0}^{n}\frac{\left(
-1\right)  ^{k}}{n-k!k!}\operatorname{Tr}[\mathcal{T}_{G}(H^{n-k})H^{k}]
\end{align}

Let us consider the term%
\begin{equation}
\sum_{k=0}^{n}\frac{\left(  -1\right)  ^{k}}{n-k!k!}\operatorname{Tr}%
[\mathcal{T}_{G}(H^{n-k})H^{k}].
\end{equation}
Suppose that $n$ is odd. Then consider, with the substitution $\ell=n-k$, that%
\begin{align}
&  \sum_{k=0}^{n}\frac{\left(  -1\right)  ^{k}}{n-k!k!}\operatorname{Tr}%
[\mathcal{T}_{G}(H^{n-k})H^{k}]\nonumber\\
&  =\sum_{k=0}^{\left(  n-1\right)  /2}\frac{\left(  -1\right)  ^{k}}%
{n-k!k!}\operatorname{Tr}[\mathcal{T}_{G}(H^{n-k})H^{k}]\nonumber\\
&  \qquad+\sum_{k=\left(  n+1\right)  /2}^{n}\frac{\left(  -1\right)  ^{k}%
}{n-k!k!}\operatorname{Tr}[\mathcal{T}_{G}(H^{n-k})H^{k}]\\
&  =\sum_{k=0}^{\left(  n-1\right)  /2}\frac{\left(  -1\right)  ^{k}}%
{n-k!k!}\operatorname{Tr}[\mathcal{T}_{G}(H^{n-k})H^{k}]\nonumber\\
&  \qquad+\sum_{\ell=0}^{\left(  n-1\right)  /2}\frac{\left(  -1\right)
^{n-\ell}}{\ell!n-\ell!}\operatorname{Tr}[\mathcal{T}_{G}(H^{\ell})H^{n-\ell
}]\\
&  =\sum_{k=0}^{\left(  n-1\right)  /2}\frac{\left(  -1\right)  ^{k}}%
{n-k!k!}\operatorname{Tr}[\mathcal{T}_{G}(H^{n-k})H^{k}]\nonumber\\
&  \qquad+\sum_{\ell=0}^{\left(  n-1\right)  /2}\frac{\left(  -1\right)
^{n-\ell}}{\ell!n-\ell!}\operatorname{Tr}[\mathcal{T}_{G}(H^{n-\ell})H^{\ell
}]\\
&  =\sum_{k=0}^{\left(  n-1\right)  /2}\frac{\left(  -1\right)  ^{k}}%
{n-k!k!}\operatorname{Tr}[\mathcal{T}_{G}(H^{n-k})H^{k}]\nonumber\\
&  \qquad+\sum_{k=0}^{\left(  n-1\right)  /2}\frac{\left(  -1\right)  ^{n-k}%
}{k!n-k!}\operatorname{Tr}[\mathcal{T}_{G}(H^{n-k})H^{k}]\\
&  =\sum_{k=0}^{\left(  n-1\right)  /2}\frac{\left(  -1\right)  ^{k}+\left(
-1\right)  ^{n-k}}{n-k!k!}\operatorname{Tr}[\mathcal{T}_{G}(H^{n-k})H^{k}]\\
&  =0.
\end{align}
The second-to-last line follows from the fact that the twirl is its own
adjoint and from cyclicity of trace. For the last line, consider that $\left(
-1\right)  ^{k}+\left(  -1\right)  ^{n-k}=0$ for all $k\in\{0,\ldots, (n-1)/2\}$ when $n$ is odd.

Suppose instead that $n$ is even. Then setting $\ell=n-k$ we find that%
\begin{align}
&  \sum_{k=0}^{n}\frac{\left(  -1\right)  ^{k}}{n-k!k!}\operatorname{Tr}%
[\mathcal{T}_{G}(H^{n-k})H^{k}]\nonumber\\
&  =\sum_{k=0}^{n/2}\frac{\left(  -1\right)  ^{k}}{n-k!k!}\operatorname{Tr}%
[\mathcal{T}_{G}(H^{n-k})H^{k}]\nonumber\\
&  \qquad+\sum_{k=n/2+1}^{n}\frac{\left(  -1\right)  ^{k}}{n-k!k!}%
\operatorname{Tr}[\mathcal{T}_{G}(H^{n-k})H^{k}]\\
&  =\sum_{k=0}^{n/2}\frac{\left(  -1\right)  ^{k}}{n-k!k!}\operatorname{Tr}%
[\mathcal{T}_{G}(H^{n-k})H^{k}]\nonumber\\
&  \qquad+\sum_{\ell=0}^{n/2-1}\frac{\left(  -1\right)  ^{n-\ell}}{\ell
!n-\ell!}\operatorname{Tr}[\mathcal{T}_{G}(H^{\ell})H^{n-\ell}]\\
&  =\sum_{k=0}^{n/2}\frac{\left(  -1\right)  ^{k}}{n-k!k!}\operatorname{Tr}%
[\mathcal{T}_{G}(H^{n-k})H^{k}]\nonumber\\
&  \qquad+\sum_{k=0}^{n/2-1}\frac{\left(  -1\right)  ^{n-k}}{k!n-k!}%
\operatorname{Tr}[\mathcal{T}_{G}(H^{k})H^{n-k}]\\
&  =\sum_{k=0}^{n/2}\frac{\left(  -1\right)  ^{k}}{n-k!k!}\operatorname{Tr}%
[\mathcal{T}_{G}(H^{n-k})H^{k}]\nonumber\\
&  \qquad+\sum_{k=0}^{n/2-1}\frac{\left(  -1\right)  ^{n-k}}{k!n-k!}%
\operatorname{Tr}[\mathcal{T}_{G}(H^{n-k})H^{k}]\\
&  =\frac{\left(  -1\right)  ^{n/2}}{\left(  n/2!\right)  ^{2}}%
\operatorname{Tr}[\mathcal{T}_{G}(H^{n/2})H^{n/2}]\nonumber\\
&  \qquad+\sum_{k=0}^{n/2-1}\frac{\left(  -1\right)  ^{k}}{n-k!k!}%
\operatorname{Tr}[\mathcal{T}_{G}(H^{n-k})H^{k}]\nonumber\\
&  \qquad+\sum_{k=0}^{n/2-1}\frac{\left(  -1\right)  ^{n-k}}{k!n-k!}%
\operatorname{Tr}[\mathcal{T}_{G}(H^{n-k})H^{k}]\\
&  =\frac{\left(  -1\right)  ^{n/2}}{\left(  n/2!\right)  ^{2}}%
\operatorname{Tr}[\mathcal{T}_{G}(H^{n/2})H^{n/2}]\\
&  \qquad+\sum_{k=0}^{n/2-1}\frac{\left(  -1\right)  ^{k}+\left(  -1\right)
^{n-k}}{n-k!k!}\operatorname{Tr}[\mathcal{T}_{G}(H^{n-k})H^{k}]\nonumber\\
&  =\frac{\left(  -1\right)  ^{n/2}}{\left(  n/2!\right)  ^{2}}%
\operatorname{Tr}[\mathcal{T}_{G}(H^{n/2})H^{n/2}]\\
&  \qquad+\sum_{k=0}^{n/2-1}\frac{2\left(  -1\right)  ^{k}}{n-k!k!}%
\operatorname{Tr}[\mathcal{T}_{G}(H^{n-k})H^{k}]\nonumber\\
&  =\sum_{k=0}^{n/2}\frac{\left(  2-\delta_{k,n/2}\right)  \left(  -1\right)
^{k}}{n-k!k!}\operatorname{Tr}[\mathcal{T}_{G}(H^{n-k})H^{k}].
\end{align}
Then the overall formula is given by%
\begin{align}
&  \frac{1}{d\left\vert G\right\vert }\sum_{g\in G}\operatorname{Tr}[U^{\dag
}(g)e^{iHt}U(g)e^{-iHt}]\nonumber\\
&  =\frac{1}{d}\sum_{n=0}^{\infty}\left(  -1\right)  ^{n}t^{2n}\times
\nonumber\\
&  \qquad\sum_{k=0}^{n}\frac{\left(  2-\delta_{k,n}\right)  \left(  -1\right)
^{k}}{2n-k!k!}\operatorname{Tr}[\mathcal{T}_{G}(H^{2n-k})H^{k}]\\
&  =\frac{1}{d}\sum_{n=0}^{\infty}\frac{\left(  -1\right)  ^{n}}{\left(
2n\right)  !}t^{2n}\times\nonumber\\
&  \qquad\sum_{k=0}^{n}\binom{2n}{k}\left(  2-\delta_{k,n}\right)  \left(
-1\right)  ^{k}\operatorname{Tr}[\mathcal{T}_{G}(H^{2n-k})H^{k}].
\end{align}

Now let us establish the expansion in \eqref{eq:acc_prob_beauty}. By applying the Baker--Campbell--Hausdorff formula and the nested commutator
in \eqref{eq:def-nested-comm}, consider that%
\begin{align}
& \operatorname{Tr}[U^{\dag}(g)e^{iHt}U(g)e^{-iHt}]\nonumber\\
& =\operatorname{Tr}\left[  U^{\dag}(g)\sum_{n=0}^{\infty}\frac{\left[
\left(  iHt\right)  ^{n},U(g)\right]  }{n!}\right]  \\
& =\sum_{n=0}^{\infty}\frac{\left(  it\right)  ^{n}}{n!}\operatorname{Tr}%
\left[  U^{\dag}(g)\left[  \left(  H\right)  ^{n},U(g)\right]  \right] 
\end{align}
As derived above, it is only necessary to consider even powers in $t$ when including the sum over $g\in G$, and so we consider the following:
\begin{multline}
\sum_{n=0}^{\infty}\frac{\left(  it\right)  ^{2n}}{2n!}\operatorname{Tr}%
\left[  U^{\dag}(g)\left[  \left(  H\right)  ^{2n},U(g)\right]  \right]  \\
 =\sum_{n=0}^{\infty}\frac{\left(  -1\right)  ^{n}t^{2n}}{2n!}%
\operatorname{Tr}\left[  U^{\dag}(g)\left[  \left(  H\right)  ^{2n}%
,U(g)\right]  \right]  .
\end{multline}
 Then we find that
\begin{widetext}
\begin{align}
 \operatorname{Tr}\left[  U^{\dag}(g)\left[  \left(  H\right)  ^{2n}%
,U(g)\right]  \right]  &
 =\operatorname{Tr}\left[  U^{\dag}(g)\left[  H,\left[  \left(  H\right)
^{2n-1},U(g)\right]  \right]  \right]  \\
& =\operatorname{Tr}\left[  U^{\dag}(g)\left(  H\left[  \left(  H\right)
^{2n-1},U(g)\right]  -\left[  \left(  H\right)  ^{2n-1},U(g)\right]  H\right)
\right]  \\
& =\operatorname{Tr}\left[  \left(  U^{\dag}(g)H-HU^{\dag}(g)\right)  \left[
\left(  H\right)  ^{2n-1},U(g)\right]  \right]  \\
& =\operatorname{Tr}\left[  \left[  U^{\dag}(g),H\right]  \left[  \left(
H\right)  ^{2n-1},U(g)\right]  \right]  \\
& =\operatorname{Tr}\left[  \left[  U^{\dag}(g),H\right]  \left(  H\left[
\left(  H\right)  ^{2n-2},U(g)\right]  -\left[  \left(  H\right)
^{2n-2},U(g)\right]  H\right)  \right]  \\
& =\operatorname{Tr}\left[  \left[  \left[  U^{\dag}(g),H\right]  ,H\right]
\left[  \left(  H\right)  ^{2n-2},U(g)\right]  \right]  \\
& =\operatorname{Tr}\left[  \left[  \left[  U^{\dag}(g),H\right]  ,H\right]
\left(  H\left[  \left(  H\right)  ^{2n-3},U(g)\right]  -\left[  \left(
H\right)  ^{2n-3},U(g)\right]  H\right)  \right]  \\
& =\operatorname{Tr}\left[  \left[  \left[  \left[  U^{\dag}(g),H\right]
,H\right]  ,H\right]  \left[  \left(  H\right)  ^{2n-3},U(g)\right]  \right]
\\
& =\operatorname{Tr}\left[  \left[  U^{\dag}(g),\left(  H\right)  ^{n}\right]
\left[  \left(  H\right)  ^{n},U(g)\right]  \right]  \\
& =\operatorname{Tr}\left[  \left(  \left[  \left(  H\right)  ^{n}%
,U(g)\right]  \right)  ^{\dag}\left[  \left(  H\right)  ^{n},U(g)\right]
\right]  \\
& =\Big\Vert \left[  \left(  H\right)  ^{n},U(g)\right]  \Big\Vert _{2}%
^{2}.
\end{align}
\end{widetext}
The third-to-last line follows from induction and the second-to-last from the
fact that%
\begin{equation}
\left[  Y^{\dag},\left(  X\right)  ^{n}\right]  ^{\dag}=\left[  \left(
X\right)  ^{n},Y\right]  ,
\label{eq:cascade-dagger}
\end{equation}
for Hermitian $X$ and by using the convention that%
\begin{align}
\left[  Y^{\dag},\left(  X\right)  ^{n}\right]    & \equiv[\cdots
\lbrack\lbrack Y^{\dag},\underbrace{X],X]\cdots,X}_{n\text{ times}}],\\
\left[  Y^{\dag},(X)^{0}\right]    & \equiv Y^{\dag}.
\end{align}
Eq.~\eqref{eq:cascade-dagger} follows from applying $[A,B]^\dag = [B^\dag,A^\dag]$ inductively.
Plugging back in above, we find that%
\begin{multline}
\frac{1}{d\left\vert G\right\vert }\sum_{g\in G}\operatorname{Tr}[U^{\dag
}(g)e^{iHt}U(g)e^{-iHt}]\\
=\sum_{n=0}^{\infty}\frac{\left(  -1\right)  ^{n}t^{2n}}{d\left\vert
G\right\vert \left(  2n!\right)  }\sum_{g\in G}\Big\Vert \left[  \left(
H\right)  ^{n},U(g)\right]  \Big\Vert _{2}^{2}.
\end{multline}

\section{Derivation of Acceptance Probability of the Second (Variational) Hamiltonian Symmetry Test}

\label{app:variational-sym-test}

Here we present an alternative derivation of \eqref{eq:modifiedacc}, as well as a derivation of
\eqref{eq:accept-prob-optimized} and \eqref{eq:lower-bnd-est-small-t}. Suppose that the input to the circuit in Figure~\ref{fig:altcircuit} is a pure state
$|\psi\rangle$, rather than the maximally mixed state. Then the initial state
of the algorithm is given by
\begin{equation}
\frac{1}{\sqrt{\left\vert G\right\vert }}\sum_{g\in G}|g\rangle_{C}
|\psi\rangle.
\end{equation}
After the first controlled unitary, the Hamiltonian evolution $e^{-iHt}$, and
the second controlled unitary, the state becomes
\begin{equation}
\frac{1}{\sqrt{\left\vert G\right\vert }}\sum_{g\in G}|g\rangle_{C}
U(g)e^{-iHt}U^{\dag}(g)|\psi\rangle.
\end{equation}
The acceptance probability is then given by
\begin{align}
& \left\Vert
\begin{array}
[c]{c}
\left(  \langle+|_{C}\otimes \mathbb{I}\right)  \times\\
\left(  \frac{1}{\sqrt{\left\vert G\right\vert }}\sum_{g\in G}|g\rangle
_{C}U(g)e^{-iHt}U^{\dag}(g)|\psi\rangle\right)
\end{array}
\right\Vert _{2}^{2}\nonumber\\
& =\left\Vert
\begin{array}
[c]{c}
\left(  \frac{1}{\sqrt{\left\vert G\right\vert }}\sum_{g^{\prime}\in G}\langle
g^{\prime}|_{C}\otimes \mathbb{I}\right)  \times\\
\left(  \frac{1}{\sqrt{\left\vert G\right\vert }}\sum_{g\in G}|g\rangle
_{C}U(g)e^{-iHt}U^{\dag}(g)|\psi\rangle\right)
\end{array}
\right\Vert _{2}^{2}\\
& =\left\Vert \frac{1}{\left\vert G\right\vert }\sum_{g^{\prime},g\in
G}\langle g^{\prime}|g\rangle_{C}U(g)e^{-iHt}U^{\dag}(g)|\psi\rangle
\right\Vert _{2}^{2}\\
& =\left\Vert \frac{1}{\left\vert G\right\vert }\sum_{g\in G}U(g)e^{-iHt}
U^{\dag}(g)|\psi\rangle\right\Vert _{2}^{2}.
\end{align}

First let us suppose that the maximally mixed state is input. This is
equivalent to picking a pure state $|\psi_{x}\rangle$ from an orthonormal
basis, with probability $1/d$. Then in this case, the acceptance probability
is given by
\begin{align}
& \frac{1}{d}\sum_{x=1}^{d}\left\Vert \frac{1}{\left\vert G\right\vert }
\sum_{g\in G}U(g)e^{-iHt}U^{\dag}(g)|\psi_{x}\rangle\right\Vert _{2}
^{2}\nonumber\\
& =\frac{1}{d}\sum_{x=1}^{d}\left(  \frac{1}{\left\vert G\right\vert }
\sum_{g^{\prime}\in G}\langle\psi_{x}|U(g^{\prime})e^{iHt}U^{\dag}(g^{\prime
})\right)  \times\nonumber\\
& \qquad\left(  \frac{1}{\left\vert G\right\vert }\sum_{g\in G}U(g)e^{-iHt}
U^{\dag}(g)|\psi_{x}\rangle\right)  \\
& =\frac{1}{d\left\vert G\right\vert ^{2}}\sum_{x=1}^{d}\sum_{g^{\prime},g\in
G}\langle\psi_{x}|U(g^{\prime})e^{iHt}U^{\dag}(g^{\prime})\times\nonumber\\
& \qquad\qquad\qquad U(g)e^{-iHt}U^{\dag}(g)|\psi_{x}\rangle\\
& =\frac{1}{d\left\vert G\right\vert ^{2}}\sum_{x=1}^{d}\sum_{g^{\prime},g\in
G}\operatorname{Tr}[U^{\dag}(g)|\psi_{x}\rangle\langle\psi_{x}|\times
\nonumber\\
& \qquad\qquad\qquad U(g^{\prime})e^{iHt}U^{\dag}(g^{\prime})U(g)e^{-iHt}]\\
& =\frac{1}{d\left\vert G\right\vert ^{2}}\sum_{g^{\prime},g\in G}
\operatorname{Tr}\left[  U^{\dag}(g)U(g^{\prime})e^{iHt}U^{\dag}(g^{\prime
})U(g)e^{-iHt}\right]  \\
& =\frac{1}{d\left\vert G\right\vert ^{2}}\sum_{g^{\prime},g\in G}
\operatorname{Tr}\left[  U^{\dag}(g^{\prime-1}\circ g)e^{iHt}U(g^{\prime
-1}\circ g)e^{-iHt}\right]  \\
& =\frac{1}{d\left\vert G\right\vert }\sum_{g\in G}\operatorname{Tr}\left[
U^{\dag}(g)e^{iHt}U(g)e^{-iHt}\right]  .
\end{align}
The second-to-last equality follows from the group property and the fact that $U(g)$ is a representation of $g$. So this provides
an alternate proof of \eqref{eq:modifiedacc}.

Now let us prove the expansion in \eqref{eq:accept-prob-fixed-state}. Consider that
\begin{align}
&  \left\Vert \frac{1}{\left\vert G\right\vert }\sum_{g\in G}U(g)e^{-iHt}%
U^{\dag}(g)|\psi\rangle\right\Vert _{2}^{2}\nonumber\\
&  =\left\Vert \mathcal{T}_{G}(e^{-iHt})|\psi\rangle\right\Vert _{2}^{2}\\
&  =\langle\psi|\mathcal{T}_{G}(e^{iHt})\mathcal{T}_{G}(e^{-iHt})|\psi
\rangle\\
&  =\langle\psi|\mathcal{T}_{G}\left(  \mathbb{I}+iHt-H^{2}t^{2}/2+O(\tau^{3})\right)
\times\nonumber\\
&  \qquad\mathcal{T}_{G}\left(  \mathbb{I}-iHt-H^{2}t^{2}/2+O(\tau^{3})\right)
|\psi\rangle\\
&  =\langle\psi|\left(  \mathbb{I}+it\mathcal{T}_{G}(H)-\left(  t^{2}/2\right)
\mathcal{T}_{G}(H^{2})+O(\tau^{3})\right)  \times\nonumber\\
&  \qquad\left(  \mathbb{I}-it\mathcal{T}_{G}(H)-\left(  t^{2}/2\right)  \mathcal{T}%
_{G}(H^{2})+O(\tau^{3})\right)  |\psi\rangle\\
&  =1+t^{2}\langle\psi|\left(  \mathcal{T}_{G}(H)\right)  ^{2}|\psi
\rangle-t^{2}\langle\psi|\mathcal{T}_{G}(H^{2})|\psi\rangle+O(\tau^{3})\\
&  =1-t^{2}\langle\psi|\left(  \mathcal{T}_{G}(H^{2})-\left(  \mathcal{T}%
_{G}(H)\right)  ^{2}\right)  |\psi\rangle+O(\tau^{3})\\
&  =1-t^{2}\left\langle \mathcal{T}_{G}(H^{2})-\left(  \mathcal{T}%
_{G}(H)\right)  ^{2}\right\rangle _{\psi}+O(\tau^{3}).
\end{align}
The Kadison--Schwarz inequality \cite[Theorem~2.3.2]{bhatia07positivedefinitematrices} implies the following operator inequality:
\begin{equation}
    \mathcal{T}_{G}(H^{2}) \geq \left(  \mathcal{T}%
_{G}(H)\right)  ^{2}.
\end{equation}
As a consequence, the following inequality holds for every state $\ket{\psi}$:
\begin{equation}
    \left\langle \mathcal{T}_{G}(H^{2})-\left(  \mathcal{T}%
_{G}(H)\right)  ^{2}\right\rangle_{\psi} \geq 0.
\end{equation}

If we perform a maximization of the acceptance probability over every
input state $|\psi\rangle$, then it is equal to
\begin{align}
& \max_{|\psi\rangle: \left\Vert \ket{\psi}\right\Vert_2=1}\left\Vert \frac{1}{\left\vert G\right\vert }\sum_{g\in
G}U(g)e^{-iHt}U^{\dag}(g)|\psi\rangle\right\Vert _{2}^{2}\nonumber\\
& =\left\Vert \frac{1}{\left\vert G\right\vert }\sum_{g\in G}U(g)e^{-iHt}
U^{\dag}(g)\right\Vert _{\infty}^{2}\\
& =\left\Vert \frac{1}{\left\vert G\right\vert }\sum_{g\in G}\left(  \left[
U(g),e^{-iHt}\right]  +e^{-iHt}U(g)\right)  U^{\dag}(g)\right\Vert _{\infty
}^{2}\\
& =\left\Vert \frac{1}{\left\vert G\right\vert }\sum_{g\in G}\left(  \left[
U(g),e^{-iHt}\right]  U^{\dag}(g)+e^{-iHt}\right)  \right\Vert _{\infty}%
^{2}\\
& =\left\Vert e^{-iHt}+\frac{1}{\left\vert G\right\vert }\sum_{g\in G}\left[
U(g),e^{-iHt}\right]  U^{\dag}(g)\right\Vert _{\infty}^{2}\\
& \geq\left(  \left\Vert e^{-iHt}\right\Vert _{\infty}-\left\Vert \frac
{1}{\left\vert G\right\vert }\sum_{g\in G}\left[  U(g),e^{-iHt}\right]
U^{\dag}(g)\right\Vert _{\infty}\right)  ^{2}\\
& =\left(  1-\left\Vert \frac{1}{\left\vert G\right\vert }\sum_{g\in G}\left[
U(g),e^{-iHt}\right]  U^{\dag}(g)\right\Vert _{\infty}\right)  ^{2}\\
& \geq\left(  1-\frac{1}{\left\vert G\right\vert }\sum_{g\in G}\left\Vert
\left[  U(g),e^{-iHt}\right]  U^{\dag}(g)\right\Vert _{\infty}\right)  ^{2}\\
& =\left(  1-\frac{1}{\left\vert G\right\vert }\sum_{g\in G}\left\Vert \left[
U(g),e^{-iHt}\right]  \right\Vert _{\infty}\right)  ^{2}\\
& \geq1-\frac{2}{\left\vert G\right\vert }\sum_{g\in G}\left\Vert \left[
U(g),e^{-iHt}\right]  \right\Vert _{\infty}.
\end{align}
The first inequality follows from the reverse triangle inequality. The next
equality follows because $\left\Vert e^{-iHt}\right\Vert _{\infty}=1$. The
second inequality follows from the triangle inequality. The final equality
follows from the unitary invariance of the spectral norm. Thus we have established \eqref{eq:accept-prob-optimized}.

Now suppose that $\left\Vert H\right\Vert _{\infty}t<1$. Then we find that%
\begin{align}
& \left\Vert \left[  U(g),e^{-iHt}\right]  \right\Vert _{\infty}\notag \\
& =\left\Vert \left[  U(g),\mathbb{I}-iHt+\sum_{n=2}^{\infty}\frac{\left(  -iHt\right)
^{n}}{n!}\right]  \right\Vert _{\infty}\\
& =\left\Vert -it\left[  U(g),H\right]  +\left[  U(g),\sum_{n=2}^{\infty}%
\frac{\left(  -iHt\right)  ^{n}}{n!}\right]  \right\Vert _{\infty}\\
& \leq t\left\Vert \left[  U(g),H\right]  \right\Vert _{\infty}+\left\Vert
\left[  U(g),\sum_{n=2}^{\infty}\frac{\left(  -iHt\right)  ^{n}}{n!}\right]
\right\Vert _{\infty}\\
& \leq t\left\Vert \left[  U(g),H\right]  \right\Vert _{\infty}+2\left\Vert
\sum_{n=2}^{\infty}\frac{\left(  -iHt\right)  ^{n}}{n!}\right\Vert _{\infty
}\\
& \leq t\left\Vert \left[  U(g),H\right]  \right\Vert _{\infty}+2\sum
_{n=2}^{\infty}\frac{\left(  \left\Vert H\right\Vert _{\infty}t\right)  ^{n}%
}{n!}\\
& \leq t\left\Vert \left[  U(g),H\right]  \right\Vert _{\infty}+2\left(
\left\Vert H\right\Vert _{\infty}t\right)  ^{2}\sum_{n=2}^{\infty}\frac{1}%
{n!}\\
& =t\left\Vert \left[  U(g),H\right]  \right\Vert _{\infty}+2\left(
\left\Vert H\right\Vert _{\infty}t\right)  ^{2}\left(  e-2\right)  \\
& \leq t\left\Vert \left[  U(g),H\right]  \right\Vert _{\infty}+2\left\Vert
H\right\Vert _{\infty}^{2}t^{2},
\end{align}
where the second-to-last inequality follows from the assumption that $\left\Vert H\right\Vert _{\infty}t<1$.
This implies that%
\begin{multline}
\frac{1}{\left\vert G\right\vert ^{2}}\max_{|\psi\rangle}\left\Vert \sum_{g\in
G}U(g)e^{-iHt}U^{\dag}(g)|\psi\rangle\right\Vert _{2}^{2}\\
\geq1-\frac{2t}{\left\vert G\right\vert }\sum_{g\in G}\left\Vert \left[
U(g),H\right]  \right\Vert _{\infty}-4\left\Vert H\right\Vert _{\infty}%
^{2}t^{2},
\end{multline}
thus establishing \eqref{eq:lower-bnd-est-small-t}.

We now prove \eqref{eq:qma-nested-comm-bnd}. Consider that%
\begin{align}
& \left\Vert \frac{1}{\left\vert G\right\vert }\sum_{g\in G}U(g)e^{-iHt}%
U^{\dag}(g)\right\Vert _{\infty}^{2}\nonumber\\
& =\left\Vert \frac{1}{\left\vert G\right\vert }\sum_{g\in G}e^{iHt}%
U(g)e^{-iHt}U^{\dag}(g)\right\Vert _{\infty}^{2}\\
& =\left\Vert \frac{1}{\left\vert G\right\vert }\sum_{g\in G}\sum
_{n=0}^{\infty}\frac{\left[  \left(  iHt\right)  ^{n},U(g)\right]  }%
{n!}U^{\dag}(g)\right\Vert _{\infty}^{2}\\
& =\left\Vert \sum_{n=0}^{\infty}\frac{\left(  it\right)  ^{n}}{n!}\frac
{1}{\left\vert G\right\vert }\sum_{g\in G}\left[  \left(  H\right)
^{n},U(g)\right]  U^{\dag}(g)\right\Vert _{\infty}^{2}\\
& =\left\Vert \mathbb{I}+\sum_{n=1}^{\infty}\frac{\left(  it\right)  ^{n}}{n!}\frac
{1}{\left\vert G\right\vert }\sum_{g\in G}\left[  \left(  H\right)
^{n},U(g)\right]  U^{\dag}(g)\right\Vert _{\infty}^{2}\\
& \geq\left(  \left\Vert \mathbb{I}\right\Vert _{\infty}-\left\Vert \sum_{n=1}^{\infty
}\frac{\left(  it\right)  ^{n}}{n!}\frac{1}{\left\vert G\right\vert }%
\sum_{g\in G}\left[  \left(  H\right)  ^{n},U(g)\right]  U^{\dag
}(g)\right\Vert _{\infty}\right)  ^{2}\\
& =\left(  1-\left\Vert \sum_{n=1}^{\infty}\frac{\left(  it\right)  ^{n}}%
{n!}\frac{1}{\left\vert G\right\vert }\sum_{g\in G}\left[  \left(  H\right)
^{n},U(g)\right]  U^{\dag}(g)\right\Vert _{\infty}\right)  ^{2}\\
& \geq\left(  1-\sum_{n=1}^{\infty}\frac{t^{n}}{n!}\frac{1}{\left\vert
G\right\vert }\sum_{g\in G}\left\Vert \left[  \left(  H\right)  ^{n}%
,U(g)\right]  U^{\dag}(g)\right\Vert _{\infty}\right)  ^{2}\\
& =\left(  1-\sum_{n=1}^{\infty}\frac{t^{n}}{n!}\frac{1}{\left\vert
G\right\vert }\sum_{g\in G}\left\Vert \left[  \left(  H\right)  ^{n}%
,U(g)\right]  \right\Vert _{\infty}\right)  ^{2}.
\end{align}
In the above, we employed unitary invariance of the spectral norm, the Baker--Campbell--Hausdorff formula, and the triangle inequality.

\section{Further Examples: Weakly-Coupled NMR and Heisenberg XY model}

\label{app:extra}

\begin{figure}
\begin{center}
\includegraphics[
width=2.7in
]{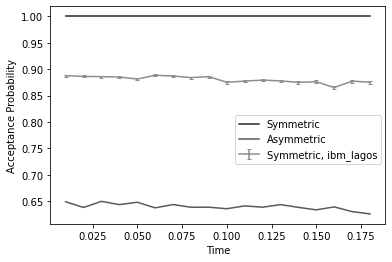}
\end{center}
\caption{Acceptance probability of our Hamiltonian symmetry testing algorithm over time, for the NMR Hamiltonian example and a group for which the Hamiltonian is either symmetric or asymmetric. The acceptance probability decays as time gets larger for the asymmetric case, even ideally. Example calculations using \textit{ibm\_lagos} show a large degree of initial symmetry before noise begins to dominate.}
\label{fig:combined}
\end{figure}

To exhibit our first algorithm, we consider two different examples in this appendix. First, we consider the example dynamics given by a weakly $J$-coupled NMR Hamiltonian \cite{van1996multidimensional}. This Hamiltonian can be expressed as 
\begin{equation}
    H_{\textrm{NMR}} \coloneqq \frac{\omega_1}{2} \sigma^Z_{1} + \frac{\omega_2}{2} \sigma^Z_{2} + \frac{\pi J}{2} \sigma^Z_{1} \otimes \sigma_{2}^Z,
\end{equation}
with units of $\hbar = 1$, where $\omega_i$ is the frequency associated to  spin $i\in\{1,2\}$ and $J$ is the coupling constant. This can be written as a diagonal matrix in the $\hat{z}$ basis; therefore, the time-evolution dynamics are also given by a diagonal matrix, as shown below. Due to this simplicity, these dynamics can be easily simulated on noisy quantum computers for appropriately short times using a two-qubit system. 

It is clear that $H_\textrm{NMR}$ is symmetric with respect to the group generated by taking the Pauli-$Z$ gates on either qubit---this corresponds to a representation of the group $\mathbb{Z}_2 \times \mathbb{Z}_2$. It is not, however, symmetric under the group generated by the CNOT and SWAP gates acting on the two-qubit system---corresponding to $D_3$, the triangular dihedral group. Thus, using these two groups as described in our algorithms, we can visualize examples of both symmetry and asymmetry, as shown in Figure \ref{fig:combined}. To generate this Hamiltonian, we define the terms $\omega_{\textrm{AVG}}=\frac{1}{2}(\omega_1 + \omega_2)$ and $\Delta\omega = \omega_2 - \omega_1$ as is common. Then the Hamiltonian can be written as:
\begin{equation}
    H_{\text{NMR}} =
    \begin{pmatrix}
-\omega_{\text{AVG}} + \frac{\pi J}{2} & 0 & 0 & 0 \\
0 & \frac{\Delta\omega-\pi J}{2} & 0 & 0\\
0 & 0 & -\frac{\Delta\omega+\pi J}{2} & 0\\
0 & 0 & 0 & \omega_{\text{AVG}} + \frac{\pi J}{2}\\
\end{pmatrix}.
\end{equation}

\begin{figure}
\begin{center}
\includegraphics[width=3.4in]{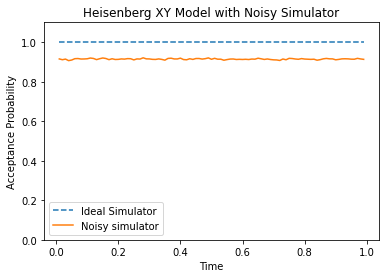}
\end{center}
\caption{Results of testing the Heisenberg XY model on four qubits using IBM Quantum's noisy simulator. The symmetries in question are given by acting simultaneously on all systems by the stated sigma matrix.}
\label{fig:HXY}
\end{figure}

As an additional example, we consider the Heisenberg XY model  \cite{LSM61}. The Hamiltonian under consideration is given by
\begin{equation}
    H_{\text{XY}} \coloneqq J \left( \sum_{i=1}^{N-1} \sigma^X_i \otimes \sigma^X_{i+1} + \sigma^Y_i \otimes \sigma^Y_{i+1}\right) \, ,
\end{equation}
where  $J > 0$ is the antiferromagnetic
exchange interaction between spins.
This Hamiltonian showcases symmetry with respect to conjugation by each Pauli matrix acting on all qubits simultaneously. That is,
\begin{align}
    [H_{\text{XY}},\sigma^X_1 \otimes \cdots \otimes \sigma^X_N] = 0 , \\
    [H_{\text{XY}},\sigma^Y_1 \otimes \cdots \otimes \sigma^Y_N] = 0 ,\\
    [H_{\text{XY}},\sigma^Z_1 \otimes \cdots \otimes \sigma^Z_N] = 0 .
\end{align}
The symmetry group to consider in this case is thus $\{\pm 1, \pm i\} \times \{I^{\otimes N}, (\sigma^X)^{\otimes N},(\sigma^Y)^{\otimes N},(\sigma^Z)^{\otimes N}\} $. As the global phase factors are irrelevant in this case, we can test this symmetry by having two control qubits prepared in a uniform superposition, one of which activates $(\sigma^X)^{\otimes N}$ and the other activating $(\sigma^Z)^{\otimes N}$. We tested this symmetry by implementing our algorithm on the IBM Quantum noisy simulator, with $N=4$, and find that the symmetry test passes with reasonable probability, as indicated in Figure~\ref{fig:HXY}. The fact that the acceptance probability is not exactly equal to one has to do with the noise involved in the simulation.

\end{document}